# Online QoS Modeling in the Cloud: A Hybrid and Adaptive Multi-Learners Approach

Tao Chen, Rami Bahsoon, Xin Yao
School of Computer Science
University of Birmingham, Birmingham, UK, B15 2TT
Email: {txc919, r.bahsoon, x.yao}@cs.bham.ac.uk

***Abstract*—Given the on-demand nature of cloud computing, managing cloud-based services requires accurate modeling for the correlation between their Quality of Service (QoS) and cloud configurations/resources. The resulted models need to cope with the dynamic fluctuation of QoS sensitivity and interference. However, existing QoS modeling in the cloud are limited in terms of both accuracy and applicability due to their static and semi-dynamic nature. In this paper, we present a fully dynamic multi-learners approach for automated and online QoS modeling in the cloud. We contribute to a hybrid learners solution, which improves accuracy while keeping model complexity adequate. To determine the inputs of QoS model at runtime, we partition the inputs space into two sub-spaces, each of which applies different symmetric uncertainty based selection techniques, and we then combine the sub-spaces results. The learners are also adaptive; they simultaneously allow several machine learning algorithms to model QoS function and dynamically select the best model for prediction on the fly. We experimentally evaluate our models using RUBiS benchmark and realistic FIFA 98 workload. The results show that our multi-learners approach is more accurate and effective in contrast to the other state-of-the-art approaches.**

## I. Introduction

Cloud-based services provided as Software-as-a-Service (SaaS) are typically running on top of the software stack within the Platform-as-a-Service (PaaS) layer [3]. They are also supported by the Virtual Machines (VM) and hardware within the Infrastructure-as-a-Service (IaaS) layer [4]. To offer elasticity under changing external environment conditions (e.g., workload, size of incoming job etc.), cloud service providers often dynamically scale various internal control knobs, which provide on-demand configuration of software (e.g., threads of service/application) and hardware resources (e.g., CPU and memory of VM) in a shared infrastructure. In this work, we refer to ***both*** control knobs and environment conditions in the cloud as ***primitives***. The key problem, which cloud/service providers face is how to manage runtime Quality of Service (QoS) by autoscaling to the best set of control values. By QoS, we refer to the non-functional attributes (e.g., throughput) experienced by the end-users of the cloud-based services. In particular, the fundamental challenge is how to link QoS with the primitives in cloud, which we address in this paper.

QoS models allow the use of primitive values as inputs and predict the likely QoS value as an output. These models can better express QoS sensitivity; by sensitivity, we are interested in *which* (e.g., are throughput and CPU correlated?), *when* (i.e., at which point in time they are correlated?) and *how* (i.e., the magnitude of primitives in correlation) the primitives correlate with QoS. An accurate QoS model in the cloud can serve as a powerful tool to reason and compare the consequences of different elastic autoscaling strategies. However, the difficulty is that the QoS sensitivity tends to fluctuate at runtime. In addition, QoS modeling in the cloud suffers from the problem of QoS interference, which we refer to scenarios where a service exhibits wide disparity in its QoS performance due to the fluctuation in primitives of co-located services on the VM and co-hosted VMs on the Physical Machine (PM) [24,15]. This is a typical consequence of resources contention. The QoS interference is uncertain in nature; because it is difficult to know when contention would occur and what the degree of such contention is. These runtime uncertainties urge the need for a ***fully dynamic*** and ***online*** modeling approach that continuously evolves itself.

The majority of the work on QoS modeling in cloud has been either static (e.g., queuing network [9]) or semi-dynamic [10-15]. The effectiveness of static approaches is restricted by their simplified assumptions on service's internal operations [10], which limits them for online QoS modeling in cloud. In addition, they do not take QoS interference into account. On the other hand, semi-dynamic approaches are either online [10-14] or offline [15]. The main difference between them is that the offline models tend to be limited in the way they cope with the unexpected changes at runtime [10]. They are called 'semi-dynamic' because both online and offline approaches focus on dynamically model the magnitude of primitives in correlation to QoS; while the selection of primitives to determine the features of models has been fixed and offline (e.g., only consider CPU and memory as inputs).

Existing semi-dynamic approaches tend to be limited because: (i) they do not consider software configuration (e.g., thread), which can interplay with the hardware provisions to influence QoS [20,24]; (ii) It is well-known that primitives selection can improve model accuracy [22], however, given the increasing variability of software configuration and application/service type in the cloud, the fixed selection requires heavy human intervention [10] and can result in considerable complexity and difficulty in its application. This is especially true when the primitives can influence the QoS in a ***dynamic*** and ***joint*** fashion: e.g., the service and its QoS tend to be sensitive to CPU only if the service *thread* and/or the CPU of co-hosted VM are set to a high value; (iii) The cloud suffers uncertain workload, VM deployment and QoS interference. Henceforth, the fixed selection of primitives can become invalid at runtime and mislead the modeling.

In previous work [8], we have proposed a fully dynamic, fine-grained and online QoS modeling approach for handling QoS sensitivity and interference. The approach combines *symmetric uncertainty* [21], an efficient measurement of data relevance for feature selection [8,22], with two alternative machine learning algorithms, which are *Auto-Regressive Moving Average with eXogenous inputs* [18] (ARMAX) and *Artificial Neural Network* [17] (ANN) to reach two formulations of the model. This solution and other existing modeling approaches [10-15] are called single learner-based as they apply single primitives selection technique and learning algorithm to model QoS. We have shown in [8] that the previously proposed approach is able to achieve better accuracy in contrast to existing semi-dynamic approaches. However, our subsequent investigations have revealed several limitations of such approach. Firstly, it ignores the QoS interference caused by VMs co-hosted on the same PM [24,15]. Secondly, the single learner approach can easily result in a QoS model with large numbers of inputs due to its over-sensitive handling in QoS interference. This will unnecessarily complicate the model and downgrade the prediction accuracy. Thirdly, we have observed that different learning algorithms (e.g., ANN and ARMAX) can be suitable only for certain QoS trends; however, a single learner approach requires the engineers to predetermine the suitable learning algorithm. This can entail manual and extensive investigation rendering it as an expensive process. Moreover, a predetermined approach does not cater for unexpected or envisioned changes in QoS at runtime.

In this paper, we propose an online QoS modeling approach to overcome the above limitations using hybrid and adaptive multi-learners. In particular, our novel contributions include:

*Firstly*, our previous work modeled QoS interferences of services co-located on a VM. This work, however, additionally factors interferences caused by VMs co-hosted on a PM.

*Secondly*, for primitives selection, we propose a hybrid multi-learners approach to determine *which* and *when* primitives correlates with the QoS on the fly. The idea is that we aim to select the relevant and useful primitives which can improve accuracy in the modeling. To this end, we partition the primitives space into two sub-spaces; the learner in each sub-space uses different primitives selection techniques based on *symmetric uncertainty* [21] and the results of the two learners are combined. Increasing the number of space partitions can introduce computational overhead, thus we used two partitions and we found out that it produces adequate accuracy.

*Thirdly*, we develop an adaptive multi-learner solution to dynamically model *how* the primitives correlates with the QoS. Precisely, multiple learners that apply different learning algorithms are used to build a bucket of models. By doing so, we aim to dynamically select the best learning algorithm and its resulted model during prediction in cloud. Particularly, we have examined three widely used learning algorithms as exemplars, these are: ANN, ARMAX and *Regression Tree* (RT) [19].

*Fourthly*, we implement our modeling approach based on an autonomic architecture in the cloud. We experimentally evaluate the approach under four commonly used QoS attributes, these are: response time, throughput, availability and reliability. In addition, we have used the well-known *RUBiS* [6] benchmark and the FIFA 98 [7] workload. The results reveal that our approach is effective and has acceptable overhead.

In the following, Section II decomposes the problem of QoS modeling and presents the model. Section III describes our architecture and overview of the approach. Section IV specifies the hybrid and adaptive multi-learners. Section V reports on experiments and evaluation. Section VI and VII present related work and conclusion respectively.

## II. Models and Problem Analysis

### A. Cloud System Model

We assume that cloud-based applications are composed of services, each has different QoS requirements and external environment changes (e.g., changes in workload). These applications and services are deployed on a cloud software stack, which can be setup using various control knobs. In addition, they are hosted on the cloud infrastructure where resources are shared via VMs. Often, multi-tiers applications and services in the cloud can have multiple replicas for load balancing purpose. Therefore we assume that each tier in a multi-tiers application, consisting of concrete services $\{S_1, S_2, \dots S_i\}$, can have multiple replicas deployed on different VMs even PMs; the replicas of a tier can also consist of the replicas of services that was originally contained by this tier. In this work, we refer to the replicas of concrete services as service-instances: the *jth* service-instance of the *ith* concrete service is denoted by $S_{ij}$. Unlike existing work [13], which focus on modeling for the entire application and VM, we aim to create fine-grained QoS models for each service-instance. In particular, the resulted models should cope with the QoS interferences at both inter-VMs and inter-services level. In addition, instead of modeling the effect of VM-level provisioning (i.e. add/remove a VM), we focus on the effect of fine-grained provisioning inside VM (e.g., CPU of a VM and/or *maxThread* of a service-instance). This would provide more flexible use of the model (e.g., vertical scaling) and has been becoming a trend in cloud. It is important to consider vertical scaling before horizontal scaling (e.g., add, remove or migrate VMs) as the former is often much more efficient than the later.

It is worth noting that, apart from the co-hosted services and co-located VMs, QoS interference can also occur due to contention on the functionally dependent services. For instance, $S_{11}$ and $S_{31}$ (both running on different PMs) can be both dependent on $S_{21}$ (e.g., a database service). This implies that $S_{11}$ and $S_{31}$ incur QoS interference. However, we discovered that in such case, the primitives of $S_{31}$ tend to be insignificant in the QoS modeling of $S_{11}$ as the same information has already been expressed by the primitives of $S_{21}$, which is also part of the invocation. As a result, we consider the co-hosted services and co-located VMs as the primary causes of QoS interference.

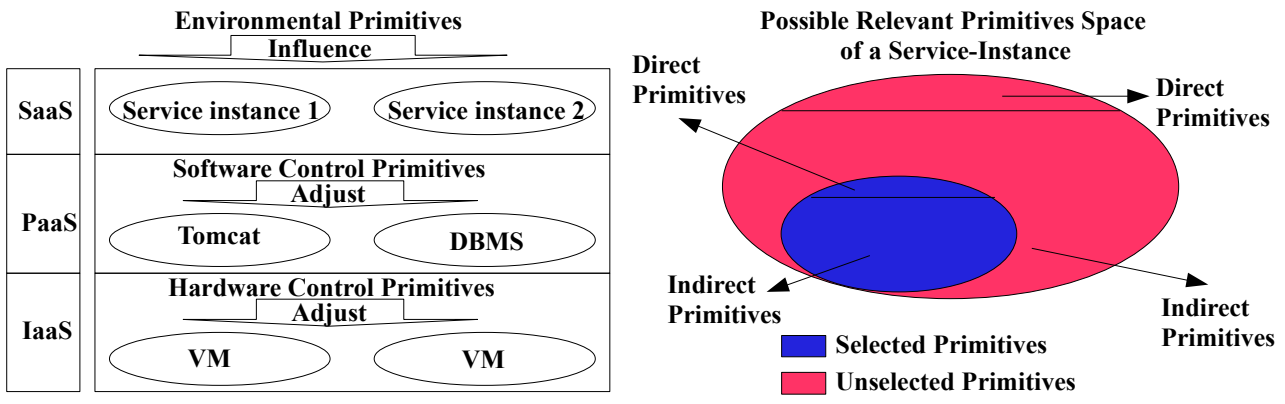

Fig. 1. The primitives decomposition (left) and partitioning (right)

### *B. Cloud Primitives Decomposition and Partitioning*

To achieve better generality, we decompose the notion of primitives into two major domains: these are ***Control Primitive (CP)*** and ***Environmental Primitive (EP)***. *Control Primitives* are the internal control knobs and can be either software or hardware, which can be managed by the cloud providers to support QoS. Specifically, software control primitives are software tactics and configurations; such as the number of threads in thread pool of service/application, the number of database connections and load balancing policies etc. Whereas, hardware control primitives are computational resources, such as CPU and memory. As shown in Figure 1 (left), software and hardware control primitives rely on the PaaS and IaaS layers respectively. In particular, it is non-trivial to consider software control primitives when modeling QoS in the cloud as they have been shown to be important features for QoS [20,24]. On the other hand, *Environmental Primitives* refer to the external stimulus that cause dynamics and uncertainty in the cloud; for examples, workload and unpredictable incoming data etc. If the cloud provider is able to control the presence of the stimulus, then these can be then considered as control primitives.

To prevent noises and improve accuracy, selecting the right primitives as inputs is critical for QoS modeling in the cloud. However, the difficulty is that the primitive inputs, which are relevant and useful for modeling QoS in the cloud tend to be dynamic [24]. All possible primitives inputs for modeling the QoS attributes of a service-instance form a space, which we call ***possible relevant primitives*** space; the problem is to select on the fly a relevant and useful subset of primitives from the space for modeling QoS. This space can be defined by:

***Rule 1:*** *A primitive belongs to the possible relevant primitives space for modeling the QoS of $S_{ab}$ if:*

*1) it is a software control or environmental primitive of $S_{ab}$, or*

*2) it is a hardware control primitive of the VM that runs $S_{ab}$, or*

*3) in case of $S_{ab}$ has direct functional dependency on $S_{cd}$, it is a software control or environmental primitive of $S_{cd}$, or*

*4) in case of $S_{ab}$ has direct functional dependency on $S_{cd}$, it is a hardware control primitive of the VM that runs $S_{cd}$, or*

*5) it is a software control or environmental primitive of $S_{cd}$, which is co-located with $S_{ab}$ on the same VM, or*

*6) it is a hardware control primitive of the VM, which is co-hosted with the VM that runs $S_{ab}$ on the same PM.*

We tackle the problem of selecting useful primitives using symmetric uncertainty [21], which measures the degree of relevance between two time series variables by producing a value ranges from 0 to 1 - a greater value implies higher relevance. At one extreme, the value between a QoS attribute and a primitive is 1 indicating that all information of the primitive is correlated with the QoS (and vice versa). At the other extreme, the value of 0 implies that changes in the primitive's behavior are independent of that of the QoS. Formally, symmetric uncertainty is calculated by:

$$U(X,Y) = \frac{2\times \sum_{y\in Y}\sum_{x\in X} p(x,y)\log\left(\frac{p(x,y)}{p(x)\times p(y)}\right)}{\sum_{x\in X} p(x)\log(p(x)) + \sum_{y\in Y} p(y)\log(p(y))} \qquad (1)$$

where $X$ and $Y$ are two series of values over time (e.g, a QoS attribute and a primitive); $x$ and $y$ are one of these values. p(x,y) is the joint probability between two values and p(x) is the marginal probability of a value. Analysis of the data of symmetric uncertainty values in [8] showed that for each feature dimension, certain primitives tend to be more relevant to the QoS than others, e.g., the CPU of the underlying VM usually has greater values than the CPU of co-hosted VMs; this fact motivates us to partition the *possible relevant primitives* space based on the relevance between primitives and QoS. By clustering on the symmetric uncertainty values of primitives; we found that using two sub-spaces can improve accuracy and reduce model complexity without having large overhead, as shown in Section V. These two sub-spaces are named: ***direct primitives*** space and ***indirect primitives*** space; the former is usually more relevant to the QoS than the later. Our aim is to select and update the subsets of primitives from these two sub-spaces on the fly, as we will see in Section IV-B. We discovered that the *direct primitives* space can be defined by:

***Rule 2:*** *A primitive belongs to the direct primitives space for modeling the QoS of $S_{ab}$ if it meets conditions 1, 2, 3 or 4 in Rule 1.*

On the other hand, the *indirect primitives* space, which mainly contains the information related to interference, is defined by:

***Rule 3:*** *A primitive belongs to the indirect primitives space for modeling the QoS of $S_{ab}$ if it meets conditions 5 or 6 in Rule 1.*

An example of the partitions is shown in Figure 1 (right). Also, start and stop VMs can trigger re-partition based on *Rule 1-3*.

### *C. QoS Model for Interference*

The generic online model for the *kth* QoS attribute of $S_{ij}$ at a given sampling interval *t* is formally expressed as:

$$QoS_k^{ij}(t) = f(SP_k^{ij}(t), \delta) \qquad (2)$$

where $QoS_k^{ij}(t)$ is the discrete or mean value (e.g., average response time) of the *kth* QoS attribute of $S_{ij}$ from *t-1* to *t*. *f* is the QoS function, which changes at runtime, as we will see in Section IV-D. $\delta$ refers to any other inputs (e.g., historical time-series QoS points and tuning variables etc) required by the algorithm to train *f* apart from the primitives. To handle QoS interferences, we denote input $SP_k^{ij}(t)$ in Eq. 2 as the ***selected primitives matrix*** of $QoS_k^{ij}(t)$ at *t*, formally depicted in Eq. 3.

$$SP_k^{ij}(t) = \begin{bmatrix} CP_a^{xy}(t) & \dots & EP_b^{mn}(t-1) & \dots \\ \dots & \dots & \dots & \dots \\ CP_a^{xy}(t-q+1) & \dots & EP_b^{mn}(t-q) & \dots \end{bmatrix} \qquad (3)$$

This matrix contains the primitive inputs of $QoS_k^{ij}(t)$ selected from the *possible relevant primitives* space for QoSs of $S_{ij}$, as we will see in Section IV-B. More concretely, the column entries indicate *which* and *when* primitives correlate with the QoS. *q* determines the number of row entries, which indicates the use of how many historical time-series points of the selected primitives as inputs can impact the learning algorithm that trains *f*. We observed that it is better to set *q* as constant for certain algorithms (e.g., *q*=1 for ANN and RT); however for the others (e.g., ARMAX), we found that *q* should be determined during training via *hill-climbing* optimization, which starts with *q=1*, then automatically increase the number of row entires one by one during training till the accuracy cannot be further improved [8]. $CP_a^{xy}(t)$ and $EP_b^{mn}(t-1)$ denote the *ath* control primitive of $S_{xy}$ and the *bth* environmental primitive of $S_{mn}$ respectively. The actual values of $CP_a^{xy}(t)$ and $EP_b^{mn}(t-1)$ that used in the modeling are the mean of measured data from *t-1* to *t* and from *t-2* to *t-1* respectively. To dynamically train the QoS function *f*, we investigate three most widely-used learning algorithms form the literature, these are ARMAX, ANN and RT. In the following, we briefly explain these learning algorithms:

*Auto-Regressive Moving Average with eXogenous inputs* - ARMAX [18] models the correlation between QoS and primitives as a linear relation and it captures the time-series information into the model. In this work, we train the ARMAX using linear Least Mean Square (LMS) approach [27]; and the *q* is determined using hill-climbing algorithm that starts with *q=1*, then automatically increase the number of row entires one by one during training till it reaches good accuracy [8].

*Artificial Neural Network* - ANN [17] is a powerful supervised learning algorithm, which is capable for modeling complex nonlinear correlations. This is achieved by weighting the inputs and transforming them using activation function to produce the output. In this work, we use three layers and Sigmoid function in the network as this setup tends to relief the issue of local minima. ANN is trained using a well-known technique, namely the Resilient backPROPagation (RPROP) [25]. We found that use *q=1* (i.e., no time series information) can produce the best result; and the right number of hidden neurons is determined using hill-climbing algorithm during training till the accuracy cannot be further improved [8].

*Regression Tree* - RT [19] is a learning algorithm that maps the relation of primitives and QoS into a tree-like structure, in which leaves represent class labels and branches express conjunctions of features to reach these labels. The tree is trained using the Classification and Regression Trees (CART) technique [26] and we found that use *q=1* (i.e., no time series information) can produce the optimal results.

To adaptively model a given $QoS_k^{ij}(t)$ online, our QoS modeling approach consists of two phases: (i) a primitives selection phase that determines the content of $SP_k^{ij}(t)$ at runtime using hybrid multi-learners; and (ii) a QoS function training phase that trains function *f* on the fly via adaptive multi-learners. Finally, the adaptive multi-learners select the best model to predict QoS.

## III. Overview of the Online QoS Modeling Approach in Cloud

As shown in Figure 2, the approach is realized as middleware using autonomic architecture with a feedback loop. The service-instances running on the VMs of a PM are managed by a dedicated *Middleware Instance* (MI), which is attached to the root domain (e.g., *Dom0* [5]) of this PM. The feedback loop runs continuously to keep the models updated.

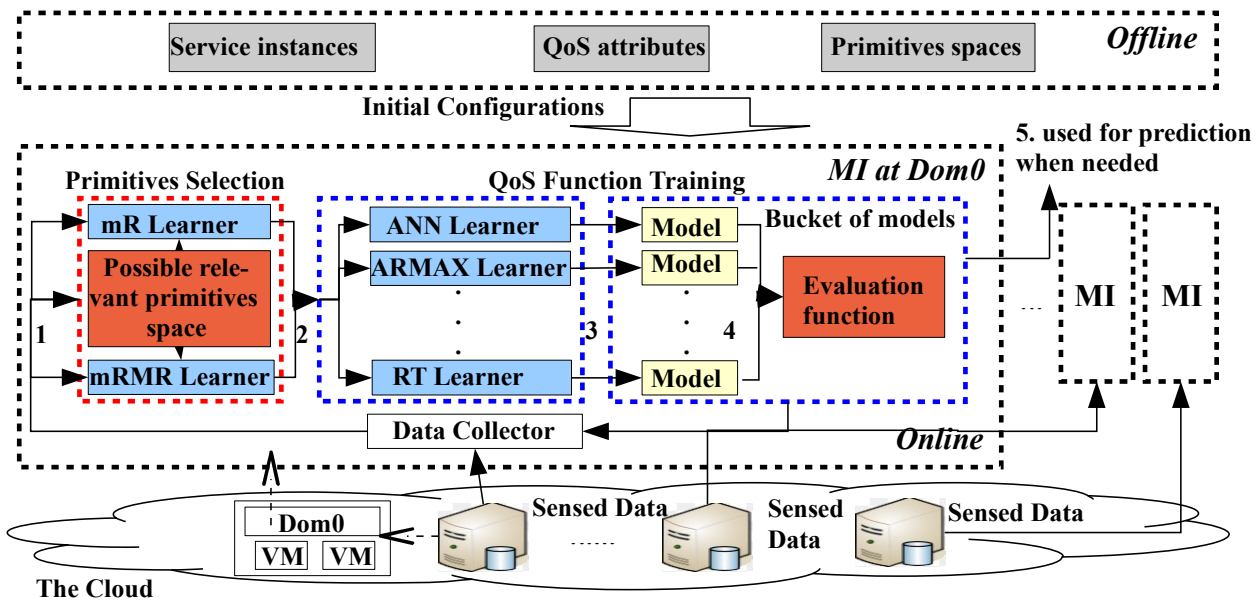

Fig. 2. Overview of the modeling approach in cloud

Our approach is designed for online scenarios; the only offline preparation is to define the current service-instances, their QoS and primitives spaces (i.e., using *Rule 1-3*). This data should be updated accordingly if changes occur. However, the approach can be also used offline in situations where conducting offline modeling in advance can further improve the online models. Within the feedback loop, *Data Collector* continuously monitors and stores sample-values of QoS and primitives from the service-instances/VMs of a PM, and those from the other PMs in the presence of functional dependency. This can be achieved by accessing the cloud sensors or log files. For each QoS attribute of a service-instance, all historical data is then passed to the primitives selection phase for determining *which* and *when* primitives correlate with QoS at runtime (step 1). Here, we have used two learners to select primitives from the *direct* and *indirect primitives* spaces (see Section IV-B). At step 2, the selected sets of primitives are combined and sent to the QoS function training phase, where multiple learners are used to model *how* the primitives correlate with QoS online (step 3). At step 4, each QoS attribute is associated with a bucket of models produced by candidate learners and an evaluation function; in addition, the weights in the evaluation function will be updated. This bucket can be then used by, e.g., an *Autoscaler* at any time (step 5). Upon prediction when given a set of inputs, the evaluation function is used to select the best model in the bucket (see Section IV-D).

## IV. Motivation and Design of the Multi-Learners

### *A. Limitation of Single Learner in Primitives Selection*

As shown in Eq. 2 and Eq. 3, to dynamically model $QoS_k^{ij}(t)$ at runtime, we first determine *which* and *when* the underlying primitives should be included as column entries in $SP_k^{ij}(t)$ for the QoS modeling. In our previous work [8], we have approached the primitives selection problem using a single learner with *maximal Relevance* (mR) technique based on symmetric uncertainty [21]. More precisely, we have used this technique to select primitives from the entire *possible relevant primitives* space for the QoSs of a given service-instance. A primitive is said to be correlated with a QoS attribute if their symmetric uncertainty value is greater than 0. We have shown that this technique produces better accuracy in contrast to the semi-dynamic models [8]. However, we have observed that a single learner based mR technique constantly results in a large number of selected primitives in the QoS model. This will result in an overcomplicated model, which can cause the following problems: firstly, it renders the decision for selecting the right elastic strategy difficult when using the model. Secondly, certain learning algorithms can easily over-fit the model and thus affect its accuracy.

To cope with this issue, we examined the data produced by the mR approach from [8] while keeping the total number of considered services and primitives unchanged. We have found that reducing the *indirect primitives* space in the modeling can improve accuracy. On the other hand, reducing the *direct primitives* space in the modeling can downgrade the accuracy as important information tends to be eliminated. This is because the *direct primitives* are able to affect QoS by directly controlling the utilization in different dimensions, which means they can provide different aspects of information; whereas all the *indirect* ones can only do so via contention, thus they can only provide information about contention and this means that the information redundancy becomes a problem. In other words, a newly selected primitive can indeed be relevant to the QoS but the same information has already been provided by another selected primitive, which has higher relevance. This fact implies that the magnitude of the new primitive is insignificant to the QoS and can downgrade the accuracy [22]. Therefore, the high redundancy can be due to over-sensitive modeling for QoS interference in the *indirect primitives* space.

### *B. Hybrid Multi-Learners for Primitives Selection*

To tackle the above issues, we design a hybrid multi-learners solution for handling the QoS interference. Precisely, instead of using all relevant primitives as inputs, we select the ones that not only relevant but also useful in the modeling. By doing so, we aim to improve accuracy and maintain adequate model complexity. The hybrid solution employs different selection techniques and produces an ensemble result online. Unlike the single learner based techniques that select primitives in the entire primitives space, we independently focus on the partitioned *direct* and *indirect primitives* space. As shown in Figure 2, for each QoS attribute of a given service-instance, we use a mR-based learner and a *maximal Relevance Minimal Redundancy* (mRMR) based learner to dynamically select useful primitives in the *direct* and *indirect primitives* space respectively. The use of mRMR is inspired by the work of [22], which applies this technique for feature selection in classification algorithms. Specifically, the objective of mR-based learner in the *direct primitives* space is to continuously select the number of primitives that maximizes Eq. 4:

$$max\,\Psi(X,y),\ \Psi=\textstyle\sum_{x\in X}^{n} U(x,y),\quad s.t.\ U(x,y)>0 \qquad (4)$$

where $x$ and $y$ are series of values of a primitive and a QoS attribute respectively. $X$ denotes the *direct primitives* space and $n$ is the number of primitives, which has been already selected. $U$ is the function of symmetric uncertainty (Eq. 1). Since the value of $U$ cannot be negative, Eq. 4 can be solved by selecting any primitives that results in non-zero value with the QoS.

On the other hand, we use a similar approach for the *indirect primitives* but with additional consideration of information redundancy. Formally, the objective of mRMR-based learner in the *indirect primitives* space is to continuously select the number of primitives that maximizes Eq. 5:

$$max\,\Phi(X,y),\Phi=\frac{\sum_{x\in X}^{n} U(x,y)}{1+\sum_{x,x'\in X} U(x,x')}\ ,s.t.\ U(x,y)>0 \qquad (5)$$

where $x'$ is also a series of concrete values of a primitive. $X$ denotes the *indirect primitives* space; and the remaining notations are the same as in Eq. 4. In this work, we apply incremental random search to optimize Eq. 5 for simplicity; however, it can be easily replaced by more sophisticated algorithms. Finally, the selected sets from both learners are combined.

**Algorithm 1** Hybrid multi-learners for primitives selection

**Inputs:** given a QoS attribute $QoS_k^{ij}$, the associated *direct primitives* space $D^{ij}$ and *indirect primitives* space $ID^{ij}$.
**Declare**: $C_{direct}$ - the collection of selected *direct primitives.*
$C_{indirect}$ - the collection of selected *indirect primitives.*
**Outputs:** the column entries of the selected primitives matrix $SP_k^{ij}(t)$

1. **start interval *t***
2. $C_{direct}:=\emptyset$, $C_{indirect}:=\emptyset$
3. $C_{direct}:=optmize\ \ \Psi(D^{ij},QoS_k^{ij})$ via Equation 4
4. $C_{indirect}:=optmize\ \ \Phi(ID^{ij},QoS_k^{ij})$ via Equation 5
5. *column entires of* $SP_k^{ij}(t):=C_{direct}\cup C_{indirect}$
6. **end interval *t***

An algorithmic description of the primitives selection phase has been illustrated in Algorithm 1. We will show (in Section V-B) that the proposed hybrid multi-learners technique leads to better accuracy as when compared to other single learner based and fixed solutions.

### *C. Limitations of Single Learner in QoS Function Training*

Recall from Eq. 2, once the primitives in $SP_k^{ij}(t)$ are selected, our next goal for QoS modeling is to determine *how* those primitives correlate with $QoS_k^{ij}(t)$ in the QoS function *f.* Previously we have evaluated two alternative machine learning algorithms, ANN and ARMAX, in the single learner to train *f* [8]. The experimental results suggest that these learning algorithms perform quite differently depending on the QoS fluctuation trends and primitives combination. This result indicates that given the generality of the proposed QoS model, the single learner is limited as we can not determine which learning algorithm to use without expensive and extensive analysis. In addition, even such process is performed, the offline analysis can still become invalid at runtime.

### *D. Adaptive Multi-Learners for QoS Function Training*

To overcome the above limitations, we propose an adaptive multi-learners technique for updating QoS function *f* on the fly and predicting the QoS values, as in Figure 2. The technique has two main processes, namely ***training*** and ***prediction***. At the training process, we ***simultaneously*** apply different learners to train the same QoS function *f*, but each of the learners uses different learning algorithm to build a model. At the prediction process, we evaluate these learning algorithms by comparing the resulted models within the bucket on the fly; the model of the best learning algorithm is used to predict QoS.

One of the most critical design decisions is to determine the evaluation function that compares the models produced by candidate learners. The basic method would be based on global mean error of all historical samples. However as shown by [10], given a set of primitive values as inputs, the most accurate model using these inputs might not be the one that has the best global error. This is because the accuracy of a model can be sensitive to the local construct of given inputs, including the variation of possible combination, scale and granularity, etc. As a result, our evaluation function aims to compare both the local error of a given inputs set produced by a model and the global error of the said model. To measure the prediction error of QoS over *n* samples, we use *Symmetric Mean Absolute Percentage Error* (SMAPE) [23], calculated as $100\times\frac{1}{n}\sum_{t=1}^{n}\frac{|predicted(t)-actual(t)|}{predicted(t)+actual(t)}$ for the total of *n* modeling intervals.

An algorithmic description of the training process has been shown in Algorithm 2. At the training process, as the collected online data increases, we continuously train two QoS models for each learner (line 2-5): (i) A *main-model* that uses 100% of the collected online data; (ii) A *sub-model*, which is trained based on 70% of the total collected data. The *sub-model* is used to test local and global error for its *main-model* of a learner. In particular, it tests the QoS prediction error against the remaining 30% testing data - the split of training and testing data follows standard machine learning approach for testing generalization errors. These generalization errors and their corresponding samples within the testing data serve as the *local error patterns* of the *main-model*. Finally, the *main-model, sub-model* and *local error patterns* are put in a bucket.

An algorithmic description of the prediction process has been shown in Algorithm 3. The prediction process is triggered when there is need to perform prediction. In particular, the best *main-model* in the bucket is used as the final model to predict QoS. To calculate the local error of a *main-model*, we leverage on the prediction error of its *sub-model* for each sample within the testing data, as recorded in the *local error patterns* (line 3-9). When given a set of inputs for predicting QoS, the local error of a *main-model* is determined by extrapolating the similarity between the given set of inputs and each sample from local error patterns; the error of the ***most similar*** sample is used as the local error (line 4-7). To this end, we apply symmetric uncertainty based *Euclidean Distance* to measure the similarity. As shown in Eq. 6, $d$ is the distance of the given set of inputs against a sample in the local error patterns.

$$d=\sqrt{\sum_{x\in X}\left(SU_x\times(P_x-P'_x)^2\right)} \qquad (6)$$

$P_x$ and $P'_x$ respectively denote the value of *xth* primitive in the given set of inputs and the value of the same primitive in a sample from local error patterns. $SU_x$ is the symmetric uncertainty value between the *xth* primitive and the QoS attribute. The sample that results in the smallest *d* is the one that we are seeking, then its corresponding error is used as the local error of the *main-model* (line 9).

**Algorithm 2** Training process in adaptive multi-learners

**Inputs:** given the column entries of $SP_k^{ij}(t)$ form *Algorithm 1* and a set of candidate learning algorithms
**Declare:** <$M_{main}$, $M_{sub}$, $L$> - a vector of *main-model*, *sub-model* and its *local error pattern.*
bucket - a collection of model vectors.
**Outputs:** a bucket of model vectors for a QoS attribute $QoS_k^{ij}$

1. **simultaneously, for each** *candidate learning algorithm* **do**
2. *find* the optimal number of row entries (i.e., the value of *q* in Eq. 3) for $SP_k^{ij}(t)$ if it has not been predefined for this learning algorithm.
3. *train main-model* $M_{main}$ and *sub-model* $M_{sub}$ *using* $SP_k^{ij}(t)$
4. *test* the *sub-model* for building *local error pattern L*
5. bucket := bucket ∪ <$M_{main}$, $M_{sub}$, $L$>
6. **end for**

**Algorithm 3** Prediction process in adaptive multi-learners

**Inputs:** given a set of inputs *P* and the bucket from *Algorithm 2*
**Declare:** $S$ - the current sample
$S_{selected}$ - the most similar sample to *P*
$d$ - the distance between *P* and the current sample
$d_{smallest}$ - the smallest distance between *P* and a sample
$E_{local}$ - the local error of the current *main-model*
$E_{global}$ - the global error of the current *main-model*
$E$ - the final error of the current *main-model*
$E_{smallest}$ - the smallest final error of a *main-model*
$M_{selected}$ - the selected *main-model* for prediction
**Outputs:** the predicted QoS value of $QoS_k^{ij}$

1. **start prediction**
2. **for each** <$M_{main}$, $M_{sub}$, $L$> in the bucket of $QoS_k^{ij}$ **do**
3. **for each** sample *S* in the *local error pattern L* of $M_{sub}$ **do**
4. *calculate* distance *d* between *P* and *S* using Equation 6
5. **if** $d_{smallest} > d$ **then**
6. $d_{smallest} := d$, $S_{selected} := S$
7. **end if**
8. **end for**
9. *get* the error of $S_{selected}$ as the local error $E_{local}$ of $M_{main}$
10. *get* the global error $E_{global}$ of $M_{main}$
11. *evaluate* final error *E* of $M_{main}$ using Equation 7
12. **if** $E_{smallest} > E$ **then**
13. $E_{smallest} := E$, $M_{selected} := M_{main}$
14. **end if**
15. **end for**
16. *predict*(*P*) using the selected *main-model* $M_{selected}$
17. **end prediction**

On the other hand, the global error of a *main-model* is the mean errors of all samples within the 30% testing data produced by its *sub-model* (line 10). Finally, the evaluation function selects the best *main-model* for a given set of inputs by examining on both the local and global error of all *main-models* in the bucket, as formally depicted in Eq. 7 (line 11-14).

$$E^i=\alpha\times E^i_{local}+\beta\times E^i_{global} \qquad (7)$$

where $E^i$, $E^i_{local}$ and $E^i_{global}$ denote the final, local and global error of the *ith main-model* respectively. $\alpha$ and $\beta$ are two heuristics expressing the relative importance of local and global errors. We have set the initial value of $\alpha$ and $\beta$ as 0.1, which means the local and global error are equally important from the beginning. The selected *main-model* and its learning algorithm for a given inputs is the one that has the smallest $E^i$ (line 16).

To capture the right weight of local and global errors, $\alpha$ and $\beta$ are updated via Eq. 8 when new data is collected.

$$\alpha=\alpha+\Delta\alpha, \beta=\beta+\Delta\beta$$
$$s.t. \begin{cases} \Delta\alpha=e_{\alpha=0,\beta=1}-e_{\alpha=1,\beta=0} & \textit{if } e_{\alpha=1,\beta=0}<e_{\alpha=0,\beta=1} \\ \Delta\beta=e_{\alpha=1,\beta=0}-e_{\alpha=0,\beta=1} & \textit{if } e_{\alpha=1,\beta=0}>e_{\alpha=0,\beta=1} \end{cases} \quad (8)$$

Specifically, $e_{\alpha=1,\beta=0}$ is the prediction error of new data produced by the selected *main-model* when $\alpha=1$ and $\beta=0$. Similarly, $e_{\alpha=0,\beta=1}$ is the error produced by the selected *main-model* when $\alpha=0$ and $\beta=1$. In this way, the error that is more useful in the selection will gradually gain more importance.

As mentioned in Section II-C, we employ three different learning algorithms (i.e., ARMAX, ANN and RT) in the adaptive multi-learners. Our approach is flexible as new algorithms can be added or old algorithms can be removed/substituted if needed. The online training of these learning algorithms follows standard machine learning procedure, interested readers could refer to our previous work [8] (ANN and ARMAX) and [14] (RT) for detailed training process in the cloud; and [17,18,19] for their detailed formulas.

## V. Expreiments and Evaluation

To evaluate our approach, we experimentally assess its accuracy and overhead. The testbed is a private cloud with numbers of PMs, each of which has Intel i7 2.8GHz Quad Cores and 4GB RAM. The PMs use Xen [5] as the hypervisor and the modeling process is running on *Dom0*. To eliminate the interference caused by modeling, we allocated one CPU core and 1.2GB RAM to *Dom0*, which tends to be sufficient. Our approach is implemented as a middleware based on Encog [2] and Apache Mathematics [1] using Java. To simulate QoS interference caused by the VMs while not exhausting resources, we run three co-hosted VMs on each PM; the remaining resources are evenly allocated to the co-hosted VMs.

Our experiments leverage on *RUBiS* [6], which is a cloud-based application consists of 26 co-located services using the *eBay.com* model. For simplicity, we have used three *RUBiS* snapshots, each of which consists of a 2-tiers (i.e., application and database tiers) based *RUBiS* application; the three *RUBiS* snapshots differ in terms of the database volume size ranging from 1GB to 5GB data. Each *RUBiS* snapshot is deployed with a software stack including Tomcat and MySQL on each co-hosted VM of a master PM; and we have implemented sensors deployed on each service-instance and VM for sending the online data to *Data Collector*. For each *RUBiS* snapshot on the master PM, the application tier is replicated to all other PMs in the cloud; and the replicated application tiers of each of the three *RUBiS* snapshots are linked to three dedicated load balancers. Three client emulators are used and they apply read/write pattern to generate requests for each load balancer.

To simulate a realistic workload within the capacity of our testbed, we vary the number of clients proportionally according to the FIFA98 workload [7], which is compressed in the way that the fluctuation of a day in the trend corresponds to 200 secs in our case. This setup can generate up to 400 parallel requests, we believe that such compression is realistic and large enough to simulate QoS interference in a public cloud. The sampling and modeling intervals are both 120 secs with the total of 500 intervals where the first 150 intervals use a static and stable workload trend aiming at providing some essential data for the modeling; whereas the rear 350 intervals follow the FIFA98 trend. This setup can generate ***one*** new sample per interval for updating the model.

TABLE I. The Examined QoS Attributes and Primitives

| **QoS and primitive** | | **Description** |
|---|---|---|
| **Output** | Response time (ms) | The leaped time between the service-instance receives and replies a request. |
| | Throughput (req/min) | The rate of completed requests. |
| | Reliability (%) | The percentage of requests that being completed less than a threshold. (30 ms) |
| | Availability (%) | The percentage of time that no requests are timeout. (60 ms) |
| **CP input** | CPU (%) | Observed CPU percentage of a VM. |
| | Memory (MB) | Observed Memory of a VM. |
| | Thread | Observed maximum concurrent threads of a service-instance. (a modified control knob of Tomcat's *maxThread* property) |
| **EP input** | Workload (req/min) | Observed request rate of a service-instance. |

### A. The QoS Attributes, Primitives and Evaluation Procedure

The concrete QoS attributes and primitives depend on scenarios. For the simplicity of exposition, we have selected commonly used QoS attributes and primitives in the evaluation, but it is worth noting that our approach is not limited to these dimensions. As listed in Table I, these QoS attributes and primitives are per-service except for CPU and memory as they are shared on a VM. For each service-instance, a QoS model can at most has 4 *direct primitives* (i.e., CPU, memory, thread and workload of the said service-instance); 54 *indirect primitives*: i.e. 2 (thread and workload)×25 (co-located service-instances)+4 (CPU and memory of another two co-hosted VMs). This combination has a size of 58 possible relevant primitives for each service-instance.

We evaluate the prediction accuracy on the fly; and for each experiment run, we examine the accuracy of one interval ahead prediction: by the end of interval *t*, the QoS models are trained based on historical data up to *t-1*, and then the predicted QoS value at *t* is compared with the actual value using SMAPE.

### B. Accuracy Results for Hybrid Multi-Learners

To assess our hybrid multi-learners (denoted as *HYBRID*) for primitives selection, we compare its effect on accuracy with three other selection techniques, these are: single learner with mR (denoted as *SINGLE-mR*), single learner with mRMR (denoted as *SINGLE-mRMR*) and the *FIXED* technique that statically uses certain primitives (CPU and memory in our case) as inputs [e.g., 10] - we modified the model from per-VM to per-service. For all cases, we apply three widely used learning algorithms (i.e., ANN, ARMAX and RT) for QoS function training under all QoS attributes. By leveraging on the procedure in Section V-A, for each selection technique, we use the SMAPE for the rear 350 out of 500 intervals in one experiment run; we calculate the mean accuracy of all service-instances on one VM of the master PM and the reported results are computed by averaging 10 runs.

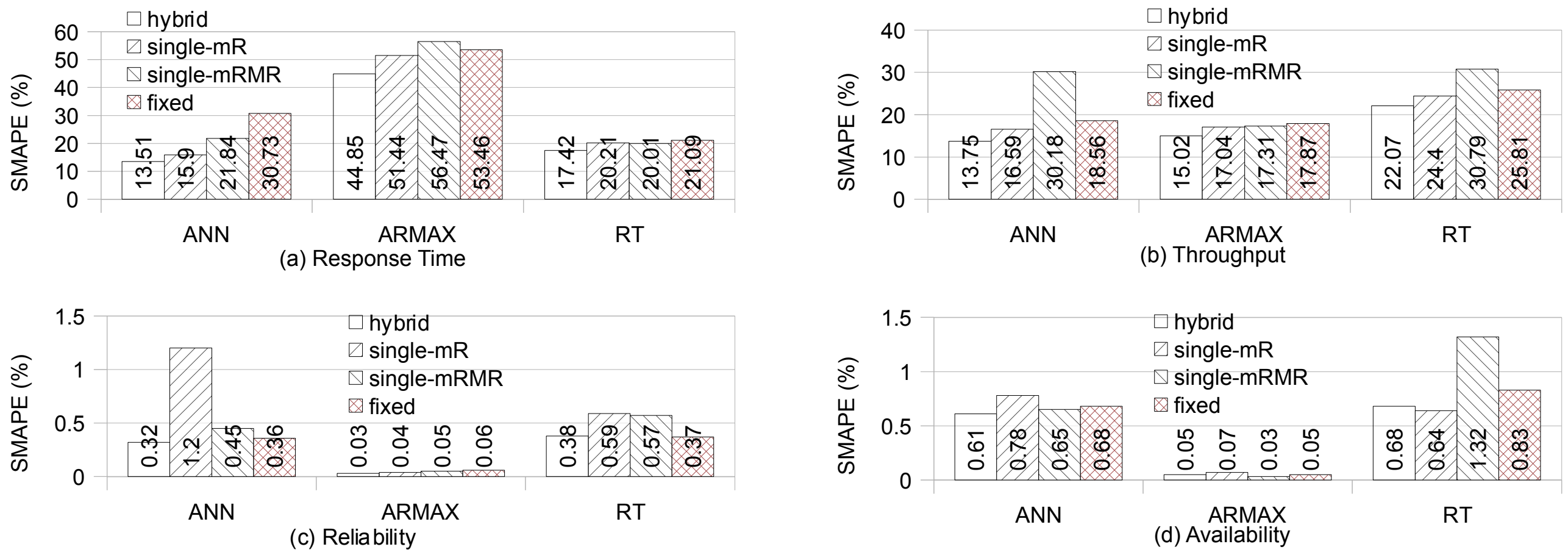

Fig. 3. Accuracy of different primitives selection techniques under various QoS attributes and learning algorithms.

TABLE II. The Number of Inputs and Mean Prediction Errors

| | **hybrid multi-learners** | **single-mR** | **single-mRMR** | **fixed** |
|---|---|---|---|---|
| **Mean SMAPE (%)** | 10.7 | 12.4 | 15 | 14.2 |
| **Number of inputs** | 5 to 8 | 30 to 44 | 2 to 3 | 2 |

The first row in Table II summaries the mean prediction error of each technique for all four QoS attributes and three learning algorithms, which we have examined; the second row shows the model complexity produced by each technique. We can see that the model produced by *hybrid* (5-8 inputs) achieves dramatic reduction on model complexity in contrast to that of *single-mR* (30-44 inputs). In addition, the *hybrid* performs better in terms of accuracy (10.7% to 12.4%). On the other hand, the *hybrid* seems to be more complex in terms of model inputs than the *single-mRMR* and *fixed*. Nevertheless, their accuracies are significantly worse than that of *hybrid* (15% and 14.2% to 10.7%); because two inputs are too simple to handle the correlation when QoS fluctuates.

To better comment on the accuracy of our technique, Figure 3(a) shows the result for response time. We can clearly see that for all three learning algorithms, applying *hybrid* in primitives selection produces the best accuracy when compared with all the single learners. In particular, the reduction on prediction error ranges from 2.39% (13.51% to 15.9% against *single-mR* on ANN) to 17.22% (13.51% to 30.73% against *fixed* on ANN). Similar result can be observed in Figure 3(b), which shows the accuracy for throughput. The *hybrid* is superior to the others for all algorithms with improvement ranging from 2.02% (15.02% to 17.04% against *single-mR* on ARMAX) to 16.43% (13.75% to 30.18% against *single-mRMR* on ANN). It is worth noting that although in most cases *hybrid* only improves on the accuracy of *single-mR* by around 2%, it can achieve such improvement with the benefits of using only 5 to 8 inputs as when compared with 30 to 44 inputs for *single-mR*. The accuracy for reliability and availability are illustrated in Figure 3(c) and 3(d). The *hybrid* is again better than the others on ANN and ARMAX for reliability; it is also the best one on ANN for availability. Interestingly, it can be noticed that there are cases where *hybrid* does not result in the best accuracy, i.e., on RT for reliability; on AMRAX and RT for availability. This is because the trends for reliability and availability are much more stable than that for response time and throughput. Therefore, the sensitivity of certain learning algorithms to the number of inputs are amplified; and this leads to over-fits. However, the differences of accuracy between *hybrid* and the best single learner ranges from 0.01% to 0.04%, which is marginal as when compared to the improvement that *hybrid* offers. In summary, our hybrid technique is able to result in adequate model complexity and reduce the prediction error, especially when QoS fluctuates.

### C. Accuracy Results for Adaptive Multi-Learners

To evaluate our adaptive multi-learners technique (denoted as *adaptive*) for QoS function training, we follow the evaluation procedure described in Section V-A. In particular, we compare the accuracy of *adaptive* with that of the single learner-based learning algorithms (i.e., ANN, ARMAX and RT) under different QoS attributes. For each learning algorithm, we examine the SMAPE for the rear 350 out of 500 intervals in each experiment run; in addition, we use the mean accuracy of all service-instances on one VM of the master PM and the reported results are the average of 10 runs. In all cases, we apply hybrid multi-learners technique for primitives selection.

In Figure 4, (a) shows the comparison result for response time: we can see that ANN is the best learning algorithm in contrast to ARMAX and RT; and the *adaptive* is able to achieve similar accuracy (13.82% error) to ANN (13.51% error). (b) illustrates the result for throughput, where the ANN again performs better than the other two. The figure clearly shows that *adaptive* (14.16% error) is only slightly worse than ANN (13.75% error). These results indicate that although the *adaptive* might occasionally produce false positive/negative for selecting the best learning algorithm, it is still able to produce very closed accuracy to the best learning algorithm for a QoS attribute. In cases of reliability (c) and availability (d), we can see that in both QoS attributes, the *adaptive* is able to produce the same prediction error (0.03% and 0.05%) as the best learning algorithm, which is ARMAX. This result means that the *adaptive* successfully determines the best learning algorithm along the QoS trend. In summary, we can note that although the algorithms behave differently depends on different QoS trends, our adaptive technique can still continuously select the suited one to predict QoS and result in good accuracy; it is also less sensitive to different QoS trends. Moreover, our solution eliminates the need of heavy human intervention and the errors that can be introduced by human analysis.

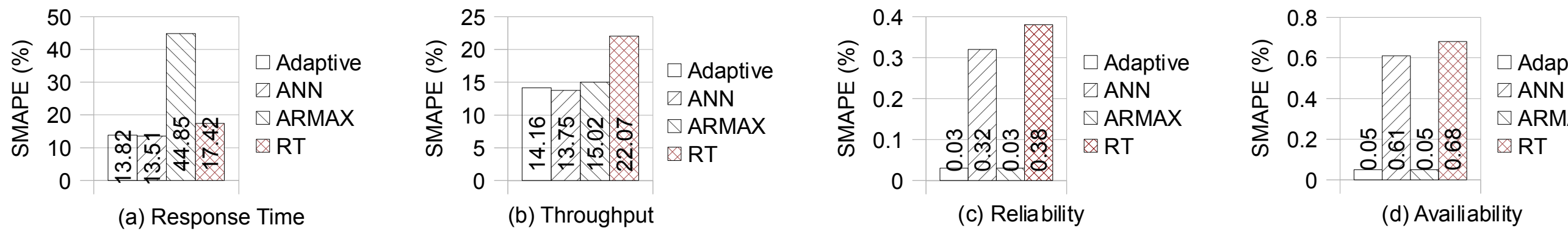

Fig. 4. Accuracy of different learning algorithms for various QoS attributes.

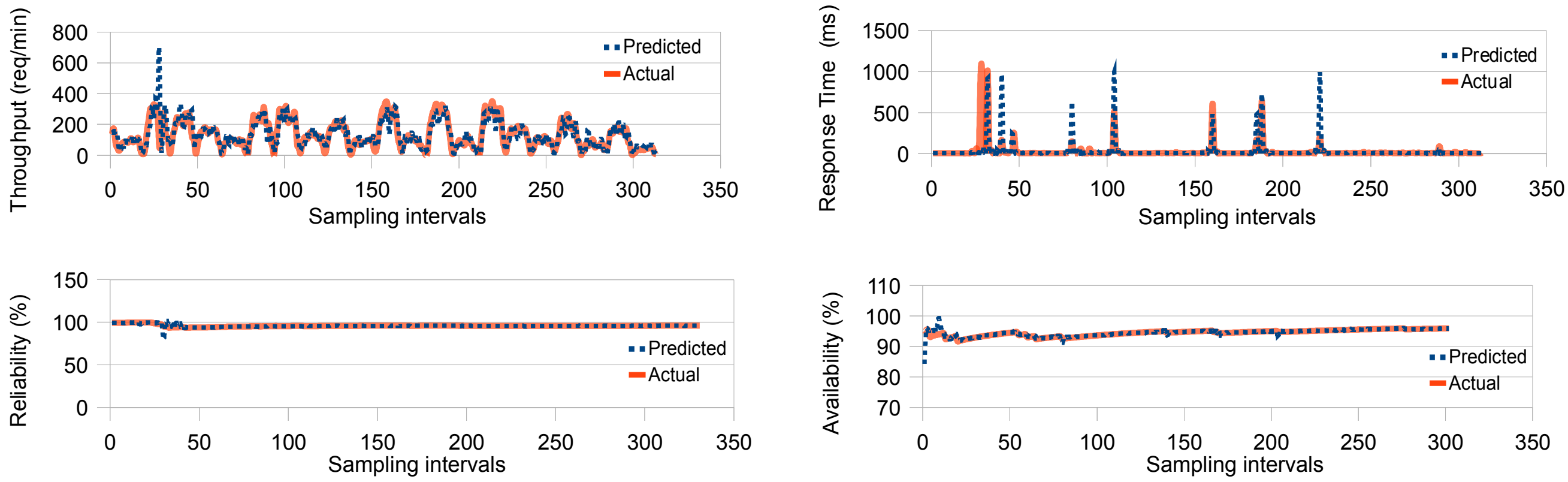

Fig. 5. The actual and predicted QoS values for an instance of the service namely *SearchItemByCategory* in one run.

### *D. Detailed Accuracy*

Figure 5 illustrates examples of the actual and predicted QoS values for all the considered QoS attributes. Due to limited space, we have used an instance of the service named *SearchItemByCategory* as the example. We can see that the prediction of the hybrid and adaptive multi-learners approach diverts from the actual QoS scale at some early peak points, e.g., the 30$^{th}$ interval for throughput. We believe that such inaccuracy is due to the applied FIFA98 trend has limited seasonality, thus the modeling approach can frequently encounter 'new behaviors' of the services at peak points, especially during the early stages of fluctuated trend. However, the figure clearly shows that the multi-learners approach is able to quickly evolve itself and detect most of the change-points in the remaining trend, given that the subsequent predictions are good even for the peak and trough.

### *E. Modeling Overhead*

To assess the overhead of our approach, we compare the execution time of *hybrid* to that of *single-mR* and *single-mRMR* for primitives selection; we also examine the execution time of *adaptive* to that of ANN, ARMAX and RT for QoS function training. We have used an instance of the service named *SearchItemByCategory* as the example and the experiments are performed using the rear 10 out of 500 intervals. We report on the average results of all QoS attributes over 10 runs.

Figure 6 (left) shows the performance overhead for different primitives selection techniques. We can see that the *hybrid* (0.65s) has relatively bigger overhead as when compared to *single-mR* (0.02s); and smaller to that of *single-mRMR* (0.84s). We have observed that this is due to the majority of overhead is caused by the optimization process of Eq 5. However, such extra overhead of *hybrid* is generally acceptable as it is still less than 1 sec. For the case of QoS function training, Figure 6 (right) illustrates the best and worst cases for all learning algorithms. In particular, ANN generally produces bigger overhead as when compared to ARMAX and RT. This is because the ANN is fundamentally more complex than the other two. For both the best and worst cases, the *adaptive* has relatively similar overhead (10s to 35.05s) to that of ANN (5.07s to 30.09s); this is expected as the *adaptive* needs to wait for the completion of all simultaneously running learning algorithms before determine the best one to use.

As a result, the overhead of our hybrid and adaptive multi-learners technique is acceptable under the sampling and modeling interval of 120s, and thus it is efficient enough to be performed online.

## VI. Related Work

The increasing complexity of managing services in the cloud makes the analytical difficulty far beyond the capability of human analysis. As a result, traditional analytical QoS model (e.g., [9]) tends to be limited in such context. To cope with this limitation, recent works have focused on single learner based and semi-dynamic approaches (e.g., ANN[10], ARMAX[11], RT [14] and linear regression[12]). Hybrid solutions are also exist: [13] adapt *kalman-filter* with linear regression to model QoS and they cluster the resulted models. However, these approaches have not considered QoS interference and dynamic selection of useful primitives. In addition, they do not take software control primitives into account thus they cannot be used in the context of PaaS.

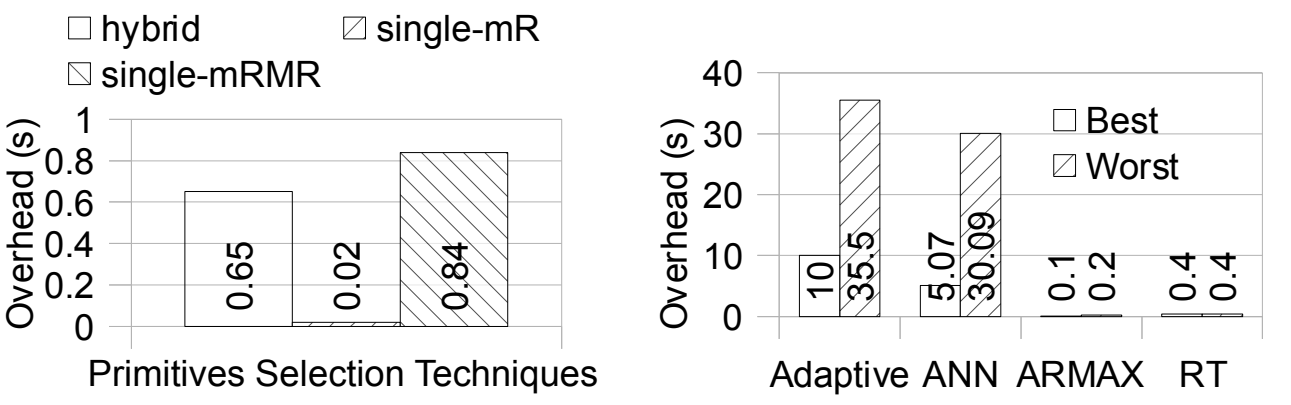

Fig. 6. Modeling overhead in terms of execution time for primitives selection (left) and QoS function training (right) on SEARCHITEMBYCATEGORY.

Despite QoS interference being core to the problem of QoS modeling in the cloud, there has been very little attempts: online [24] and offline [15] work exist for modeling QoS interference caused by the co-hosted VMs. Unlike our work, [15] do not consider software control primitives and they rely on fixed selection of primitives. [24] is the closest work to this paper as they consider software control primitives and use *Simplex Reduction* to do dynamic primitives selection, the QoS function training is handled by reinforcement learning. In contrast, our approach works on a finer model granularity of service instances rather than VMs. By doing so, we can also cope with the interferences caused by co-located services on a VM. In addition, we use hybrid and adaptive multi-learners to select the model inputs and train QoS function on the fly, which has been shown to be effective and time efficient.

Recent approaches [9,10,11,12,14] are single learner-based, which are realized through adapting a single technique for modeling *how* primitives correlate with QoS. These approaches require extensive human analysis and investigation. [16] alternatively propose a method to predict the utilization of hardware control primitive using an ensemble solution where the results from different learning algorithms are combined in a weighted-sum relation. As a result, their approach is highly sensitive to the similarity of candidate learners. Our work on the other hand, dynamically select the best algorithm for predicting the correlation between QoS and its primitives.

## VII. Conclusion and Future Work

We have proposed a novel hybrid and adaptive multi-learners approach for online QoS modeling in the cloud. In particular, we have applied a hybrid approach to determine *which* and *when* primitives correlate with QoS. We have described an adaptive solution to model *how* primitives correlate with QoS, and dynamically select the best learning algorithm for prediction. Experimentally, we have evaluated our approach with respect to accuracy and overhead using *RUBiS* and the FIFA workload. The results reveal that the proposed approach produces better and more stable accuracy than the other state-of-the-art models with adequate complexity. In addition, it results in acceptable overhead and is able to eliminate the possible errors introduced by the human for selecting modeling inputs and learning algorithms. In future work, we will report on use of the modeling approach for elastic autoscaling in the cloud.